\documentclass[reprint,twocolumn,prb,amsmath,amssymb,aps]{revtex4-2} % PRB reprint format

\usepackage{graphicx}  % For including figures
\usepackage{dcolumn}   % Align table columns on decimal points
\usepackage{bm}        % Bold math symbols
\usepackage{hyperref}  % Hyperlinks
\hypersetup{colorlinks=true, linkcolor=blue!95!black!85!yellow, citecolor=blue!95!black!85!yellow, urlcolor=blue!95!black!85!yellow} % Hyperlink settings
\usepackage{physics}   % Physics notation
\usepackage{newtxtext} % Times-like text font
\usepackage[cmintegrals]{newtxmath} % Math font
\usepackage{booktabs}  % Professional tables
\usepackage{multirow}  % Multi-row tables
\usepackage{xcolor}    % Colored text (optional)
\usepackage{float}     % Improved figure placement

\usepackage{tikz}
\newcommand{\orcidicon}{%
    \begin{tikzpicture}
        \draw[lime, fill=lime] (0,0) circle [radius=0.16] node[white] {{\fontfamily{qag}\selectfont \tiny ID}};
        \draw[white, fill=white] (-0.0625,0.006) circle [radius=0.007];
    \end{tikzpicture}
    \hspace{-2.9mm}
}
\foreach \x in {A, B}{\expandafter\xdef\csname orcid\x\endcsname{\noexpand\href{https://orcid.org/\csname orcidauthor\x\endcsname}{\noexpand\orcidicon}}}

\newcommand{\iitmPhys}{School of Physical Sciences, Indian Institute of Technology Mandi, Kamand 175075, India}
\newcommand{\iitmMech}{School of Mechanical and Materials Engineering, Indian Institute of Technology Mandi, Kamand 175075, India}

\preprint{APS/123-QED}

\begin{document}

\title{ Spin-dependent orbital selectivity and partial Kondo-screening in \\magnetically ordered Hund's metal. }

\author{Shivani Bhardwaj\orcidA{}}
\email{sbhardwajiitm369@gmail.com}
\affiliation{\iitmPhys}
\author{Sudhir K. Pandey \orcidB{}}
\email{sudhir@iitmandi.ac.in}
\affiliation{\iitmMech}

\begin{abstract}

Hund's metallicity in 3$d$ transition metal oxides constitutes a rare class of compounds, since they have been long understood considering the dominance of Hubbard $U$. $\mathrm{LiV_2O_4}$\& $\mathrm{Sr_2CoO_4}$ belong to this rare class of metals; among them, $\mathrm{LiV_2O_4}$ has been the subject of extensive investigations for its unconventional heavy-fermion behavior, while studies on $\mathrm{Sr_2CoO_4}$ remain limited despite its anomalous ferromagnetic ground state. In this study, we report an unusual spin–orbital selective localization in $\mathrm{Sr_2CoO_4}$ leading to a sharp Kondo resonance at $\sim$70 K in the spin-$up$ channel of orbitals of $t_{2g}$ symmetry using a combination of Density functional theory and Dynamical mean field theory (DFT+DMFT) calculations. Correspondingly, an appreciable reduction in the magnetization below $T$=100 K further suggests partial Kondo screening of local moments active at low temperatures, explaining its effective spin magnetization state and upturn in its resistivity observed in experimental reports. We note a significant effect of Hund's induced spin-orbital selective incoherence in dictating the temperature evolution of its macroscopic observables e.g. spin-spin correlation function and effective local moment. Our results reveal a potentially distinct/new form of spin-dependent selectivity induced via Hund's coupling in addition to the conventional orbital-selectivity in the Hund's metals, as a plausible key mechanism in stabilizing their long-range magnetic order.

%The highest figure of merit ($zT$) value of $\sim$1.05 (0.78) at 900 K with an electron doping of 10$^{18}$ (10$^{19}$) cm$^{-3}$ for LiZnAs (ScAgC) using MRTA increases significantly to $\sim$1.53 (1.0) for a 20 nm nanostructure. 

% where these properties are significantly influenced by the EPI scattering phenomena. 

%The lattice thermal conductivity obtained using phonon-phonon interaction are reduced by the nanostructuring technique at several grain sizes, which overall influences the figure of merit for these hH materials.    

%Half-Heusler (hH) compounds are currently an important research direction in the thermoelectric (TE) field due to their rare combination of low thermal conductivity and relatively favorable Seebeck coefficient and electrical conductivity.
    
\end{abstract}

\maketitle

%******************************************************** Introduction ************************************************************

%\tableofcontents
\textit{Introduction}. Hund's metals have recently emerged as a distinct class of strongly correlated systems primarily characterized by multi-orbital physics, where electronic correlations are dominated by Hund's exchange coupling ($J$) rather than the conventional direct Coulomb repulsion/Hubbard parameter ($U$)\cite{Me,L,An}. Unlike conventional metals, Hund's metals exhibit the coexistence of sizable local moments and itinerant electrons\cite{i,o,u}, leading to a rich variety of exotic correlated phases such as enhanced effective electronic masses and correlations despite large charge fluctuations,  orbital-selective correlations, spin-orbital separation, non-Fermi liquid (NFL) behavior, bad metal behavior with unusually high resistivity, etc.\cite{Kat,p,1,2,q,3,tr}.\\
These systems historically identified with iron-based superconductors\cite{Fe1} include several transition metal oxides such as ruthenates\cite{Sr,Ru}, Nickelates (e.g., $\mathrm{SrNiO_2}$, $\mathrm{LaNiO_2}$\cite{Ni,Ni2}) and intermetallic systems e.g. iron pnictides and chalcogenides\cite{Fe2,Fe3,Fe4,Mn,we}, etc.
In systems such as the iron chalcogenides (e.g., FeSe and FeTe), $J$ induced correlation effects account for the observed orbital-selective Mott transitions and NFL behavior\cite{Se,Se2}. Similarly, in both doped and undoped ruthenates (e.g., $\mathrm{Sr_2RuO_4}$, $\mathrm{Ca_2RuO_4}$, $\mathrm{Ca_{2x}Sr_xRuO_4}$\cite{Sr1,Ca,CaRu}), Hund's coupling underlies the emergence of large spin-orbital separation, leading to suppressed coherence scales and strong electronic mass renormalization.\\
The interplay between Hund's coupling, orbital-differentiation, spin/orbital screening, etc. is seen to give rise to a largely diverse range of magnetic ground states in Hund's metals. For instance, the iron-based pnictides display antiferromagnetic (AFM) spin-density-wave order, while stripe-like magnetism is noted for iron-chalcogenides\cite{T}. The ruthenates reveal Hund's driven two-stage coherence and transition between paramagnetic and AFM states depending on the structural tuning\cite{CaR,RaC}. The double-pervoskite Hund's systems such as $\mathrm{Sr_2FeMoO_6}$ exhibit ferrimagnetism with nearly full spin polarization stabilized by double-exchange\cite{O6,SrMo}. Also, the inter-metallic Hund's metals broaden the range by displaying an intriguing magnetic state, such as in helimagnetic MnSi\cite{Mn,we}.
In this diversity of rich magnetism offered, long-range magnetic ordering especially ferromagnetism appears a rarity considering most of these metals are predominantly paramagnetic, AFM or display a complex mix of them and indirect exchange interactions. 

Moreover, majority of the well-established Hund's-metal transition-metal oxides are 4$d$ electron systems, where the spatially extended $d$ orbitals result in moderate $U$ and large bandwidths, yet manage to show pronounced  correlation effects driven by $J$. Surprisingly, in striking contrast with the typical notion that the electronic correlations in 3$d$ systems are primarily driven by $U$- due to the more localized nature of 3$d$ electrons compared to 4$d$ orbitals, certain parent (undoped) 3$d$ transition-metal based oxides— e.g. $\mathrm{LiV_2O_4}$\cite{Li,V}, $\mathrm{Sr_2CoO_4}$\cite{sh} have been reported to show Hund’s metal physics. The observation of unconventional heavy-fermion behavior in magnetically frustrated $\mathrm{LiV_2O_4}$ together with the enhancement of Kondo screening and orbital-selective Mottness, highlights how Hund's coupling driven correlations can give rise of complex emergent electronic phases in even 3$d$-transition metal oxides\cite{V}. A recently proposed $\mathrm{Sr_2CoO_4}$- a 3$d$ transition metal oxide Hund’s metal, stabilizes in a long-range ferromagnetic ground state with appreciable $T_c$$\sim$250 K, which also classifies it as anomalous within its family of K$_2$NiF$4$-type compounds\cite{K2}. \\
 In this Letter, we report an unusual spin–orbital selective incoherence in $\mathrm{Sr_2CoO_4}$, a feature not previously observed in Hund’s metals, as long-range ferromagnetic ordering in such systems has not been reported before. This incoherence drives a novel spin–orbital selective Kondo-like screening of local moments at low temperatures. Interestingly, these results not only provide a microscopic explanation for the experimentally observed anomalous magnetic and transport properties of $\mathrm{Sr_2CoO_4}$, which had long remained unaddressed, but also offer insight into a potentially new form of spin-dependent selectivity induced by Hund’s coupling, in addition to conventional orbital selectivity observed in Hund's metals. 
Moreover, $\mathrm{Sr_2CoO_4}$ shows an intricate interplay between Hund’s-induced spin-orbital selective (SOS) incoherence and hybridization-mediated partial Kondo-like screening, giving rise to a rich SOS ferromagnetic phase that shares a similarity with the Kondo-lattice systems.\\

\textit{Methods}.
We have carried out fully charge self-consistent spin-polarized DFT+DMFT calculations for $\mathrm{Sr_2CoO_4}$ using the EDMFTF code\cite{edmft} using the rotationally invariant type of Coulomb interaction parametrization (full type). Here, augmented plane wave plus local orbitals
(APW+lo) method is used to carry out the DFT calculations using the WIEN2K package\cite{wien} with the Perdew-
Burke-Ernzerhof exchange functional for solids (PBEsol)\cite{pbe}.
The structural parameters of $\mathrm{Sr_2CoO_4}$ in this work are taken from our previous first-principles study along with the set of Coulomb interaction parameters $U$(4.4 eV) \& $J$ (1.16 eV)\cite{sh}.
A large hybridization energy window from -20 to 20 eV with respect to the Fermi-level is chosen. The energy convergence criterion used is $10^{-4}$ Ry/cell for 12 × 12 × 12 k-mesh size. Continuous-time quantum Monte Carlo (QMC) impurity solver is used with the ‘exacty’ double-counting (DC) scheme as it is identified to work well for metallic systems\cite{dc}. For analytical continuation of the imaginary-time self-energy, maximum-entropy method is used, as implemented in the EDMFTF code. 
%----------------------------------------

\setlength{\parindent}{3em}

\textit{Results}.
Fig.~1 shows the temperature dependence of inverse local spin susceptibility ($\chi^{-1}_{\mathrm{loc}}(T)$) and $T \chi_{\mathrm{loc}}(T)$. $\chi^{-1}_{\mathrm{loc}}$ shows appreciably linear temperature dependence in accordance with the Curie-Weiss (CW) behavior, extending to temperatures well above calculated $T_c$ ($\sim 300$~K) in the paramagnetic regime, indicating the presence of well-defined local moments above $T_c$. The CW fit to the high-temperature region yields an effective moment of $\mu_{\mathrm{eff}} = 3\, \mu_B$ corresponding to an effective spin state of $S = 1.25$. 
This value is quite underestimated considering the experimental value of $\mu_{\mathrm{eff}} = 3.67\, \mu_B$; however, it should be noted that the experimental value is obtained in the range of temperatures above only slightly higher than the $T_c$, where we expect that increasing the temperatures to further high values can lead to a different value of the effective moment.
\begin{figure}[h]
\includegraphics[width=\columnwidth]{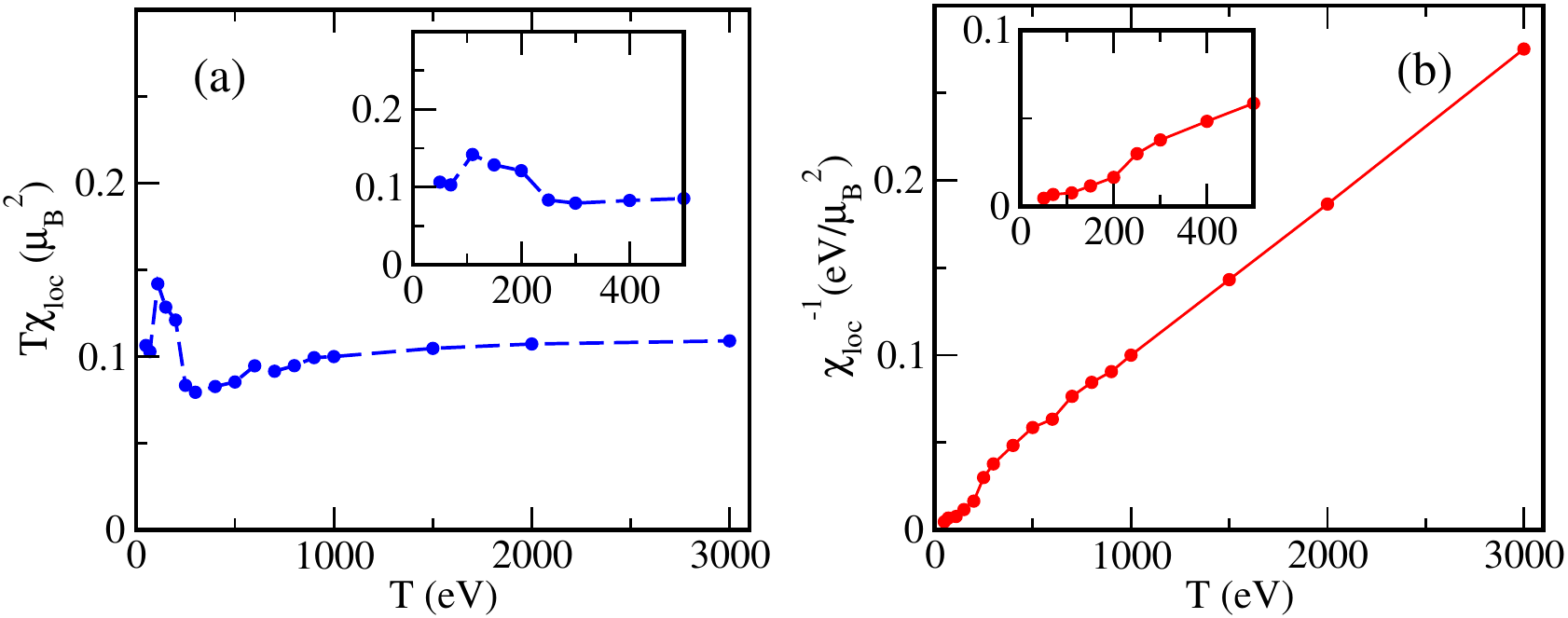} 
\caption{(a) The plot of $T \chi_{\text{loc}}$ and (b) inverse local spin-susceptibility ($\chi_{\text{loc}}^{-1}$) in the temperature range 50 -- 3000 K..}
\label{fig:f1}
\end{figure}
Below $T_c$, $\chi^{-1}_{\mathrm{loc}}$ deviates from the CW behavior, showing a crossover to Curie-like response down to 110~K, while saturates upon further lowering temperature to 70--50~K, indicative of Pauli-like behavior. Also, the presence of CW behavior is indicated in the paramagnetic regime from the 
presence of a large saturation plateau of the $T\chi_{\mathrm{loc}}$ curve above $T_c$. The upturn in $T_c$, $T\chi_{\mathrm{loc}}$ below $T_c$ corresponds to the onset of ferromagnetic fluctuations.  However, below 110~K, the $T \chi_{\mathrm{loc}}(T)$ shows a sudden dip in values between 110--50~K, signaling a departure from the monotonically increasing behavior as expected in the FM phase.\\
\begin{figure}[h]
\includegraphics[width=\columnwidth]{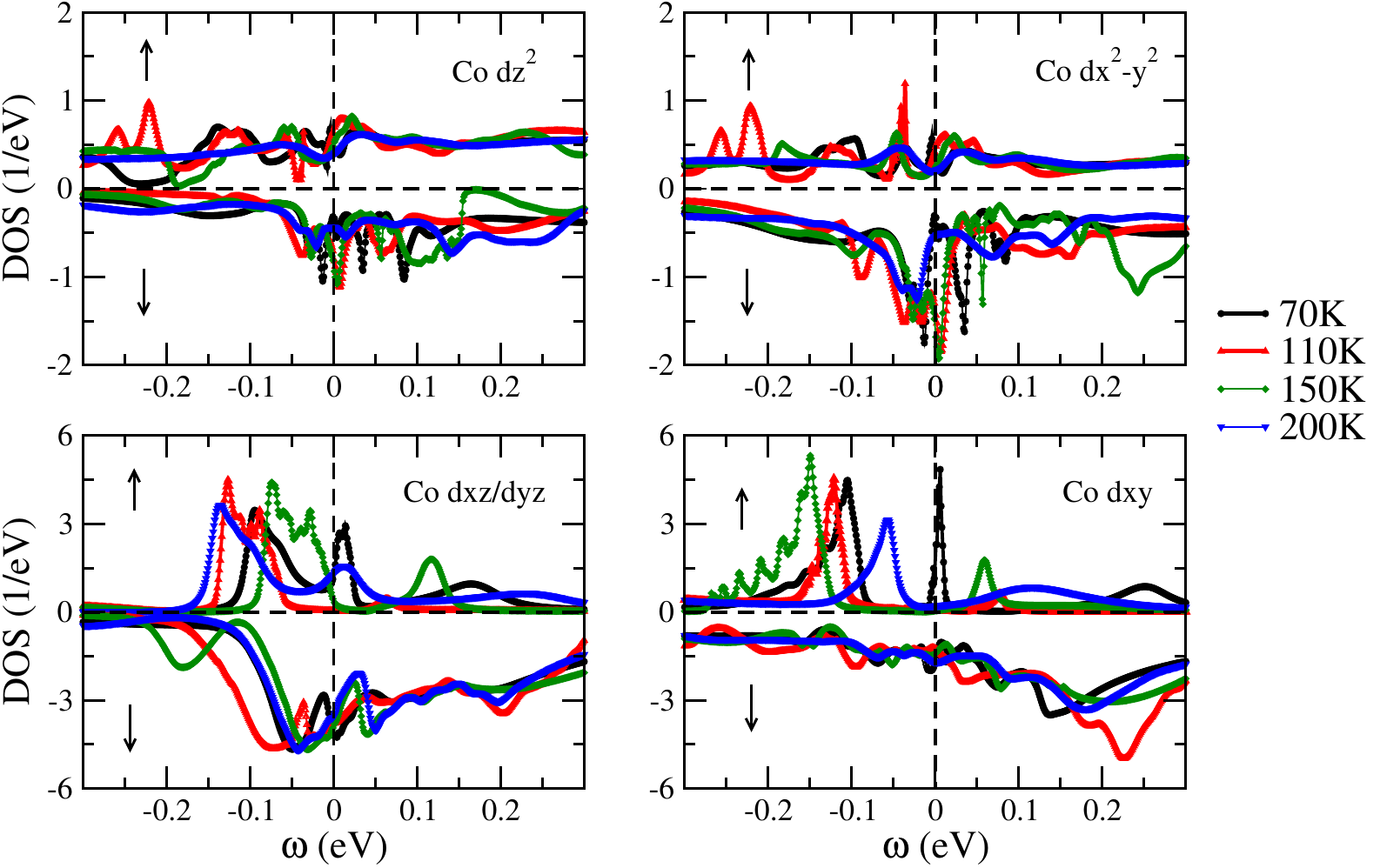} 
\caption{Temperature dependent variation of spin-orbital resolved spectral density plot for Co 3$d$ orbitals.}
\label{fig:f2}
\end{figure}
\begin{figure}[h]
\includegraphics[width=1\columnwidth]{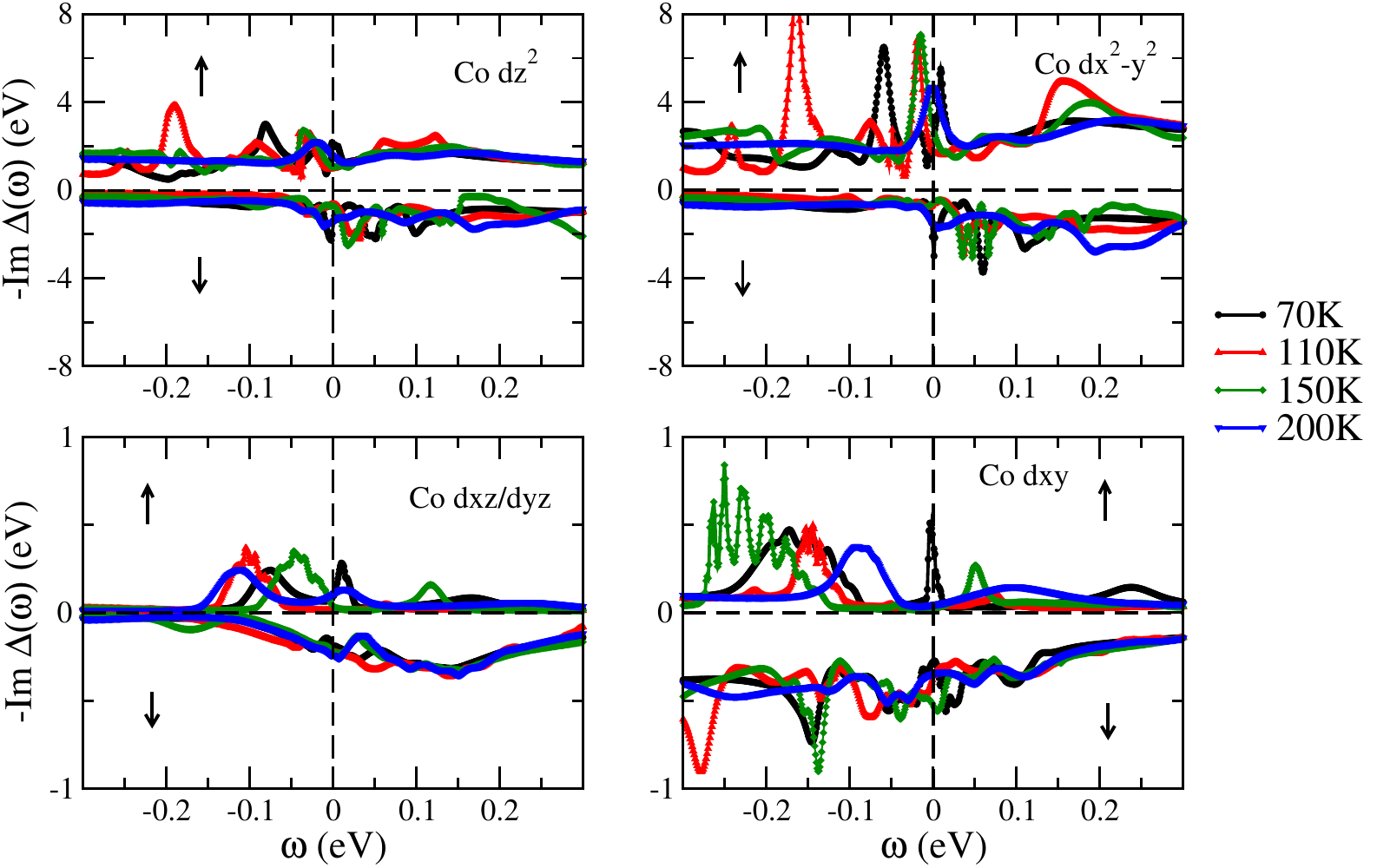} 
\caption{Temperature dependent variation of spin-orbital resolved hybridization function ($\mathrm{Im} \Delta(\omega)$) of Co 3$d$ orbitals.}
\label{fig:f3}
\end{figure}
\begin{figure*}[h]
\includegraphics[width=2\columnwidth]{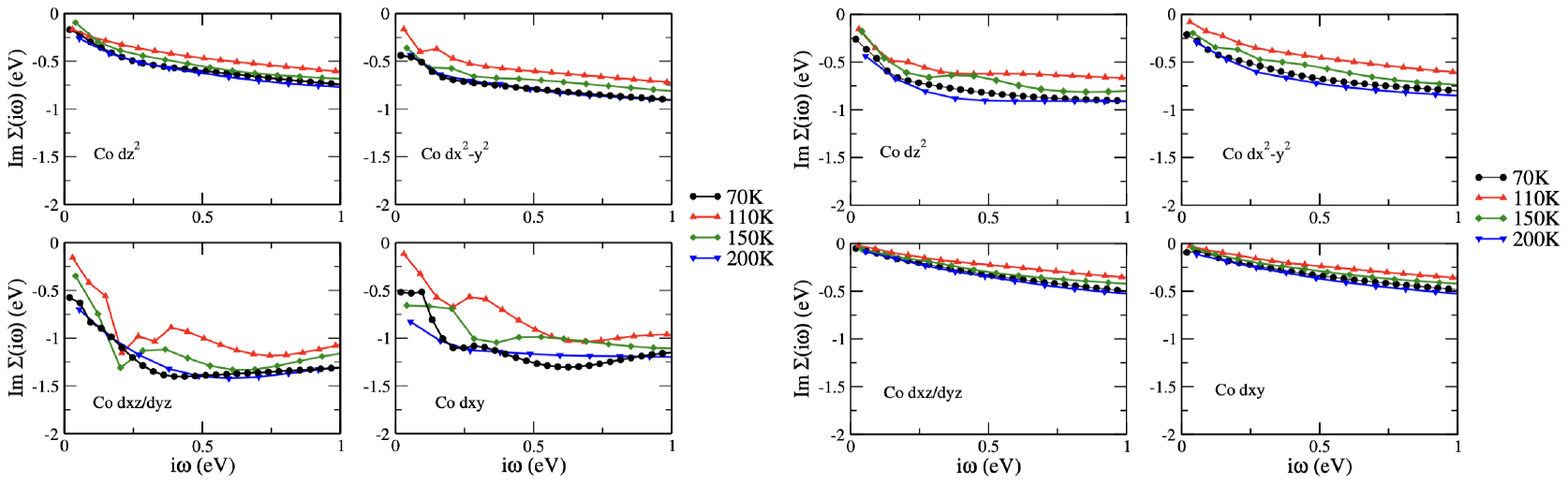} 
\caption{Temperature dependent variation of inverse quasiparticle lifetime (imaginary part of self-energy, $\mathrm{Im} \Sigma(i \omega)$) of Co 3$d$ orbitals for spin-up (Left
) and spin-down (right) channel.}
\label{fig:f4}
\end{figure*}
Next we study the temperature-dependent magnetization ($M(T)$) of SCO (refer Fig.~5 in Supplemental Material). From the M(T) curve, we note that the magnetization collapses at $T = \sim 300$~K. While decreasing the temperature from 300~K to 110~K, the $M(T)$ curve rises steadily and achieves a maximum of $2.14 \, \mu_B$ ($S$=1), which is significantly lower than the effective $S$= $1.25$ obtained from the CW fit of $\chi^{-1}_{\mathrm{loc}}(T)$, thus pointing to the onset of local moment screening active around $T_c$. Interestingly, on further lowering the temperature below 110~K, the magnetization value decreases abruptly to $\sim 1.67 \, \mu_B$ at 70~K and shows no further decrease in value till 50~K. The decrease in the $M(T)$ curve below 110~K coincides with the above noted anomalous decrease in the $T \chi_{\mathrm{loc}}(T)$ and Pauli-like behavior of 
$\chi^{-1}_{\mathrm{loc}}(T)$ below 110~K - suggesting a crossover to a partially screened local moment regime below 110~K.\\
In order to gain further insights into the microscopic origin of the temperature-dependent local moment physics, we study the temperature-dependent PDOS of the Co 3d orbitals in the FM phase (70-200~K). Fig.~2 shows the temperature-dependent evolution of the partial density of states (PDOS) of Co-3d orbitals for both spin-up ($\uparrow$) and spin-down 
($\downarrow$) channels in the vicinity of the Fermi level ($E_F$). The results reveal a distinct spin- and orbital-differentiation across the studied temperature range below the calculated magnetic transition temperature ($T_c = 300$~K). \\
In the intermediate-temperature regime (110--150~K), the spin-up channel of the $t_{2g}$ orbitals---namely $d^{\uparrow}_{xz/yz}$ and $d^{\uparrow}_{xy}$---develops a pronounced gap in their PDOS. However, at 150~K the gap in $d^{\uparrow}_{xz/yz}$ states is found to be slightly shifted above $E_F$. On increasing the temperature to 200~K, the gap closes in the 
$d^{\uparrow}_{xz/yz}$ orbitals, while the $d^{\uparrow}_{xy}$ orbital retains finite but strongly suppressed spectral weight at $E_F$. 
In contrast, the $e_g$ orbitals remain metallic in both spin channels throughout the entire temperature range, emphasizing the relative itinerancy of the $e_g$ manifold compared to the more localized $t_{2g}$ states. We also note that the PDOS of the spin-down channel exhibits very weak temperature dependence.\\
The emergence of a gap in the $t_{2g}^\uparrow$ orbitals is accompanied by a simultaneously vanishing hybridization function ($\mathrm{Im}\,\Delta(\omega)$) (see $\mathrm{Im}\,\Delta(\omega)$ of $t^{\uparrow}_{2g}$ states in Fig.~3). This dip in the charge self-consistently converged $\mathrm{Im}\,\Delta(\omega)$ arises from the feedback of dynamical electronic correlations-modified impurity, thereby reflecting the formation of robust local moments in the system with lowered effective screening. The presence of strong electronic correlations driven renormalization is further supported by the absence of such a dip in the unconverged $\mathrm{Im}\,\Delta(\omega)$ of $t^{\uparrow}_{2g}$ states obtained from the initial iterations (refer Fig.~6 in Supplemental Material). Taken together, these results point to a spin- and orbital-selective Mott-like localization within the $t_{2g}$ manifold, which stabilizes local moments in this channel thus effectively majorly contributing to the formation of local moments in the system. Such an orbital differentiation or multiorbital nature of correlation effects, even in the presence of moderate $U$, is a conventional hallmark of Hund’s metals and is very recently proposed in SCO.\\

Note that at 200~K, a pronounced yet modestly thermally broadened quasiparticle peak emerges at $E_F$ in $d_{xy}^\uparrow$ states - signaling the onset of orbital selective local moment screening around $T_c$. This onset of local moment screening by the coherent quasiparticles appearing at $T$= 200~K helps in understanding the suppressed maximum saturation magnetization achieved at 110~K, than  otherwise expected from the high temperature behavior ($S$=1.25). On the other hand, towards lower temperature scales below 110~K, a sharp coherent quasiparticle peak appears in the spin-up channel of especially $t_{2g}$ manifold. Interestingly, the width of the resonance peak is found to be of the order of its temperature energy scale (i.e.~70~K). This resonance peak width and temperature correspondence is typical of the Kondo-like resonance. Here, the PDOS behavior suggests the presence of spin-orbital selective (SOS) Kondo resonance in the system at low temperatures manifesting as partial Kondo-like screening of the local moments.\\
Concurrently, the metallic $e_g$ orbitals also exhibit the development of narrow resonance peaks below 110~K in their spin-up channel, indicating their undeniable participation in the spin-selective Kondo screening. Thus, $T = 70$~K can be identified as the onset of Kondo-screening temperature ($T_K$) for the spin-selective Kondo physics in this system.
%------------------------------------
Therefore, the results of PDOS help in understanding the distinct low- and high-temperature behaviors (flanked by the intermediate temperature regime: 110--150~K) in the macroscopic magnetic observables discussed above. The presence of anomalous temperature dependence of both $\chi^{-1}_{\mathrm{loc}}(T)$ and $T \chi_{\mathrm{loc}}(T)$ down to $T_K$ can be explained as driven by SOS selective Kondo screening leading to Pauli-like saturation in the former and reduced effective local moment in the latter. \\
Furthermore, the scattering rate given by the imaginary part of the self-energy in the Matsubara frequency domain ($\mathrm{Im}\,\Sigma(i\omega)$) (Fig.~4) also marks distinct temperature dependencies across the FM phase. $\mathrm{Im}\,\Sigma(i\omega)$ are noted  to vanish linearly for both the spin channels of the $e_g$ orbitals in the intermediate temperature range (110-150~K). Whereas, the $t^{\uparrow}_{2g}$ orbitals show largely divergent and non-vanishing variation of $\mathrm{Im}\,\Sigma(i\omega)$ towards approaching the low frequency region between 110-150~K indicating strong deviation from FL behavior. The divergent behavior of the scattering rate exhibited by $t^{\uparrow}_{2g}$ orbitals reconciles with the Mott-like gap observed in their PDOS in the corresponding temperature range.\\
Further, among the $e_g$ orbitals, the $d^{\uparrow}_{z^2}$ orbital exhibits a relatively small scattering rate at all temperatures, while the $d^{\uparrow}_{x^2 - y^2}$ orbital shows a slightly higher scattering rate, still well below the magnitude of the $t^{\uparrow}_{2g}$ orbitals. The $\mathrm{Im}\,\Sigma(i\omega)$ of $t^{\downarrow}_{2g}$ orbitals shows a remarkable Fermi-liquid-like behavior across the studied temperatures, while the scattering rate exhibited by $e^{\downarrow}_g$ orbitals is similar to their spin-up counterparts. The analysis of scattering rate provides a refined picture of the presence of partial local moments below 110~K, especially in the spin-up sector. This can be inferred from the non-analytical frequency dependence of $\mathrm{Im}\,\Sigma(i\omega)$ of the $t^{\uparrow}_{2g}$ orbitals at 70~K and also at 200~K, despite the emergence of well-defined quasiparticle peaks at $E_f$. The NFL behavior here results from the presence of partially screened local moments which introduce additional incoherent scattering channels.\\
These findings not only provide advancement in the microscopic understanding of SCO, but also reveal a striking consistency with the experimental observations that have long remained unexplained or partially understood. The central example is the long-standing puzzle of SCO's anomalously reduced observed saturation magnetization compared to the expected value from the effective local moment obtained from the CW fit in its paramagnetic phase\cite{K2,wang,Pan,sh1}. Our study shows that this reduction should naturally follow from the low-temperature SOS Kondo resonance building in the system, which partially screens the local moment. Notably, the resistivity data for SCO also offers another compelling case, as this aspect has not received enough attention in the existing literature. The experimentally observed upturn in resistivity below $\sim$100~K\cite{K2} can be accounted for, considering the active Kondo-like screening in the system at low temperatures. We also find evidence of high-temperature local moment screening observed in the present study around $\sim$200~K, correspondingly manifesting as a change in slope of the experimental resistivity curve while approaching the $T_c$.\\
Moreover, the observed finite magnetization in this system at low temperatures, despite the evidence of partial Kondo-like screening, suggests that magnetic ordering originates from the Ruderman–Kittel–Kasuya–Yosida (RKKY) type of indirect exchange interactions. The coexistence of partial Kondo screening and long-range ferromagnetic order indicates an incomplete compensation of local moments, where the unscreened local moments couple via RKKY exchange coupling to lead to a lowered spin magnetization ground state. However, it can be more rigorously accounted for via a theoretical framework describing the competition between the Kondo effect and RKKY interactions\cite{hew}. In this regard, a Kondo-lattice model treatment would be promising, as it can capture the interplay between partial Kondo screening and RKKY-driven ordering at low temperatures, thereby providing a realistic understanding of the mechanism underlying the observed lowered spin state ferromagnetism in this system. Such a similarity to Kondo-lattice systems has been reported previously for a weakly ferromagnetic Hund's metal, MnSi\cite{Mn} This behavior is surprisingly contrasting because such a competition between the Kondo screening and RKKY interactions is traditionally seen in strongly localized f-electron systems but rarely encountered in 3$d$ complexes. The identification of Hund's metal characteristics along with a significantly large effective local moment ($S$= 3/2) reported in this system plausibly causes it to mimic the strong-local moment picture analogous to f-electron systems. Such emergence of heavy-fermion behavior has also been noted in few of the 3$d$ TMOs  \cite{Hf1,Hf2}. Unlike conventional heavy-fermion systems where strong coupling with the conduction electrons and localized f-electrons drives a full Kondo singlet formation, the weaker effective coupling with the conduction electrons and Hund's coupling induced SOS localization in the present 3$d$ system likely results in only partial Kondo screening. Thereby, fostering conditions under which Kondo-like screening and RKKY-type exchange interactions can emerge and compete at low temperatures.\\
Furthermore, it is also to be noted that the single-site DMFT approach typically overestimates the magnetic transition temperatures pertaining to the lack of account of non-local correlations including the inter-site exchange interactions, which are crucial for stabilizing the long-range magnetic order in real systems. Interestingly, as opposed to this general tendency, we obtain $T_c$ ($\sim$ 300 K) in remarkably good quantitative agreement with the experimental $T_c$ (250 K)\cite{K2}. This largely suppressed $T_c$ obtained with single-site DMFT, which lies only slightly above the experimental value, suggests the primarily predominant role of Hund's coupling and local moment physics in dictating the system's $T_c$.
Conclusively, based on the results above, the electronic structure properties, in particular, the magnetic and transport properties of this system, are expected to demonstrate an intricate interplay of Hund's coupling and Kondo-like screening accompanied by SOS correlations, thereby offering a comprehensive explanation for its anomalous ferromagnetic behavior.\\  
Finally, the observed temperature dependence of $\chi_{\text{loc}}$ and magnetization, together with the analysis of experimental features, shows dominant microscopic influence from the spin-up channel of Co 3$d$ orbitals. This stems especially from the orbitals of $t_{2g}$ symmetry. As a result, spin-up channel is expected to largely control the temperature evolution of macroscopic observables such as $\chi_{\text{loc}}$, magnetization, and other low-temperature properties, establishing a Hund's coupling induced spin-polarized nature of correlation effects in the system.\\

\textit{Conclusions}.
Our results reveal three distinct temperature regimes characterizing this system: (i) a low-temperature region (below 110 K) characterized by an onset of SOS Kondo-like screening at $T_K \approx 70$ K, which accounts for the sudden drop in magnetization observed in the $M(T)$ curve below 110~K.(ii) an intermediate temperature regime ( 110–200~K) showing a spin–orbital selective localization with an effective local moment of $S$=1.(iii) a higher temperature regime with an effective $S$=1.25 spin state, above the $T_c$.
We note Hund's coupling induced unconventional SOS localization in this system as the underlying cause of the formation of large effective local moments, subsequently leading to partial Kondo-like screening at low temperatures. This further explains its anomalously lowered effective saturation magnetization and upturn in resistivity data.
Our results provide an interesting insight into studying the Hund's metals exhibiting long-range magnetic ordering considering the RKKY-type of indirect exchange mechanism in the presence of competing Kondo-effect, via Kondo-lattice models.\\

\textit{Acknowledgments}
We acknowledge the computational support provided by the High-Performance Computing (HPC) PARAM Himalaya at the Indian Institute of Technology Mandi.

%******************************************************** Methodology ************************************************************

%***************************************** Result and discussion *************************************************

\section*{Supplementary material}

\subsection{Temperature-dependent magnetization}

\begin{figure}[h]
\includegraphics[width=0.5\columnwidth]{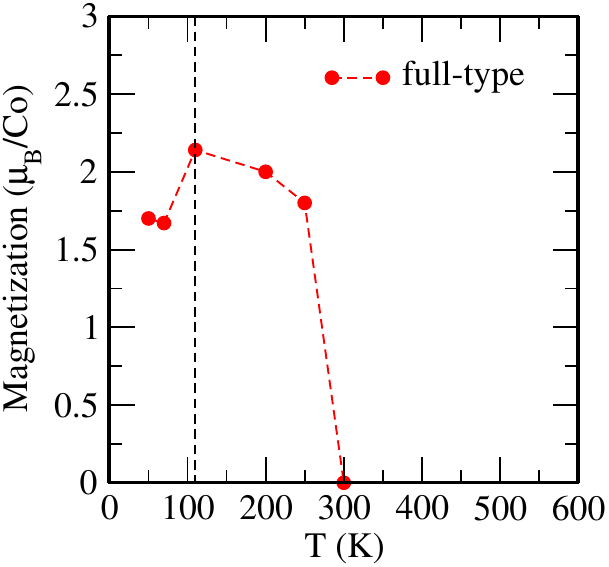} 
\caption{Temperature-dependent magnetization curve using rotationally-invariant form of Coulomb interaction parametrization (full-type) obtained from DFT+DMFT calculations.}
\label{fig:f2}
\end{figure}
\subsection{Temperature-dependent hybridization function}
\begin{figure}[h]
\includegraphics[width=1\columnwidth]{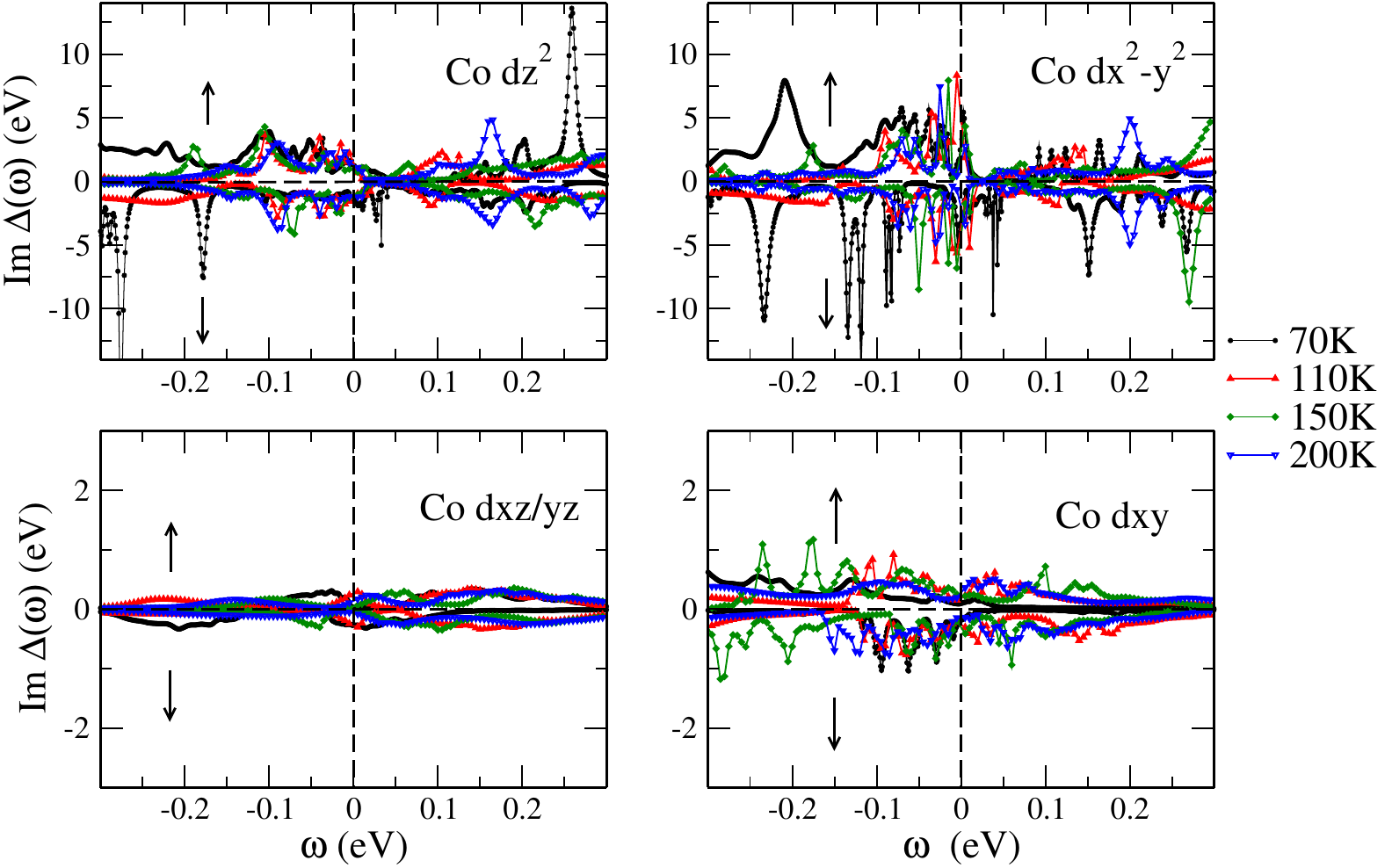} 
\caption{Temperature dependent variation of spin-orbital resolved unconverged hybridization function ($\mathrm{Im} \Delta(\omega)$) of Co 3$d$ orbitals. Note: Here, unconverged hybridization function is obtained from initial two iterations of charge self-consistent loop of DFT+DMFT calculations.}
\label{fig:f2}
\end{figure}

\end{document}